\documentclass[twocolumn,prl,showpacs,amsmath,amssymb]{revtex4}
\usepackage{graphicx}
\usepackage{bm}

\begin{document}
\title{Impurity Conduction and Magnetic Polarons in Antiferromagnetic Oxides}
\author{C. Chiorescu}
\affiliation{Department of Physics, University of Miami, Coral
Gables, Florida 33124}
\author{J.~L.~Cohn}
\affiliation{Department of Physics, University of Miami, Coral
Gables, Florida 33124}
\author{J. J. Neumeier}
\affiliation{Department of Physics, Montana State University,
Bozeman, Montana 59717}

\begin{abstract}
Low-temperature transport and magnetization measurements for the
antiferromagnets SrMnO$_3$ and CaMnO$_3$ identify an impurity band
of mobile states separated by energy $\delta$ from electrons bound
in Coulombic potentials. Very weak electric fields are sufficient
to excite bound electrons to the impurity band, increasing the
mobile carrier concentration by more than three orders of
magnitude. The data argue against the formation of self-trapped magnetic polarons
(MPs) predicted by theory, and rather imply that {\it bound} MPs
become stable only for $k_BT\ll\delta$.

\end{abstract}
 \maketitle

An electron in a magnetic solid can
perturb local moments via exchange interactions between its
spin and those of the ions, forming a self-trapped or bound magnetic polaron (MP).
Though these concepts were formulated long ago \cite{MPTheory},
experimental understanding and theoretical development of MP
physics have been limited by the relatively small number of materials found to
manifest MPs.  More recently, renewed interest in the MP problem
has been stimulated by studies of carrier-doped antiferromagnetic
(AF) manganites \cite{Reviews} and dilute magnetic semiconducting
oxides \cite{DMSOxides}.  An important emerging issue for
both classes of compounds is the energy position of donor levels
(e.g.,~associated with oxygen vacancies and/or impurities) within
the band gap and the contribution of donor-bound charge to MP
formation.

Perhaps the simplest AF systems for examining such issues are
the nominally Mn$^{4+}$ compounds, CaMnO$_3$ (CMO) and SrMnO$_3$
(SMO) which have a bipartite (G-type) AF ground state and are free
from the complex collective
interactions of Jahn-Teller-active Mn$^{3+}$ ions that
characterize more widely studied LaMnO$_3$.
They are model systems for MP studies since the
couplings between electronic, lattice, and spin degrees of freedom
for light electron doping are known \cite{ChenAllen,Meskine}.
Magnetization \cite{NeumeierCohn} and scattering \cite{Granado} studies
imply the existence of MPs in the ground state of CMO when electron doped
with La, and theory \cite{ChenAllen,Meskine} predicts these electrons
form self-trapped MPs, i.e. those bound solely by magnetic exchange interactions with ionic
spins.  However, shallow impurity states associated with oxygen vacancies
are ubiquitous in oxides, and their influence on the energetics of MP formation has received
little attention.

Here we report low-temperature transport and magnetic studies on
CMO and SMO which reveal surprising features of the donor
electronic structure that offer new insight into MP formation in
oxides.  These compounds are naturally electron doped by low
levels of oxygen vacancies ($n\sim 10^{18}-10^{19}$~cm$^{-3}$) so
that the conventional picture of donor-bound electrons in
small-radius states predicts insulating behavior at low
temperatures. Instead, we find that the low-$T$ transport involves
an impurity band of mobile electronic states separated by energy
$\delta$ from electrons bound in Coulombic potentials.  Very weak
electric fields ($F\leq 50$~V/cm) are sufficient to excite bound
electrons to the impurity band, increasing the mobile carrier
concentration by more than three orders of magnitude. The data
argue against the presence of self-trapped MPs in these compounds
and we attribute the onset of a FM contribution to the low-T
magnetization of SMO \cite{Chmaissem} to {\it bound\/} MPs which
become stable only for $k_BT\ll\delta$. The observations suggest
that the tunable mobile carrier density characteristic of the
present compounds might be observed and exploited in related
compounds for novel studies of correlated electrons.

The synthesis procedures for the single-crystal of CaMnO$_3$ (CMO)
and polycrystalline specimen of SrMnO$_3$ (SMO) are described
elsewhere \cite{NeumeierCohn,CohnPolarons,Chmaissem,CSMOeps}. The
net electron densities determined from Hall measurements at room
temperature were $n_H\equiv N_D-N_A\simeq 6\times
10^{18}$~cm$^{-3}$ (CMO) and $3\times 10^{18}$~cm$^{-3}$ (SMO).
These correspond to $\approx 10^{-3}-10^{-4}$ electrons per
formula unit, attributable to a very small oxygen deficiency.
Compensating impurities (e.g. from ppm levels in the starting chemicals) are
typical in oxides; the $T=300$~K thermopowers \cite{unpublished} of
-420~$\mu$V/K (CMO) and -180~$\mu$V/K indicate a greater
compensation in SMO.  The dc resistivity and Hall coefficient were
measured in a 9-T magnet on 6-probe specimens with silver paint
contacts. Magnetic field was applied perpendicular to the plane of
plate-like specimens in which dc currents (5~nA to 10~mA) were
applied.  Both current and field reversal were employed for Hall
and magnetoresistance measurements. A thermocouple attached to
each specimen monitored temperature rises due to self-heating,
observed only at the highest currents employed.  Magnetization
measurements were performed in a Quantum Design PPMS.
\begin{figure}
\includegraphics[width = 3.2in,clip]{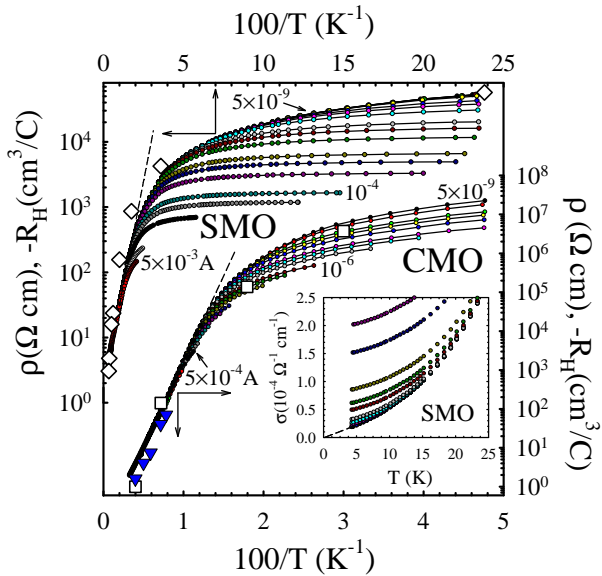}%
\vglue -.1in \caption{(Color online) $\rho$ {\it vs.}~$100/T$ for
SMO (upper abscissa, left ordinate) and CMO (lower
abscissa, right ordinate). Each curve was measured with a
different dc current, some of which are labeled; increases are in
steps of 1, 3, 5 per decade. Dashed lines are linear
fits. Hall coefficients (open squares and diamonds)
were measured at the lowest currents.  Inverted triangles
are Hall data for polycrystalline CMO from
Ref.~\onlinecite{CohnPolarons}. Inset: $\sigma(T)$ for
SMO at various currents (some omitted for clarity).} \label{RofT}
\vglue -.2in
\end{figure}

CMO and SMO are orthorhombic and cubic, respectively,
with \cite{Wollan} $T_N\simeq 125$ and \cite{Chmaissem} 230~K.  They are
weakly ferromagnetic (FM) \cite{NeumeierCohn,Chmaissem} at low $T$, with saturation
magnetizations $\sim 0.02-0.03\mu_B/$~Mn ion, but the FM moment
develops abruptly at $T<T_N$ for CMO and gradually
\cite{Chmaissem} at $T\lesssim 80$~K$\ll T_N$ for SMO.  The latter
behavior appears to be associated with FM polarons \cite{Granado},
whereas the former may be attributed to an additional FM
contribution in CMO from Dzyaloshinsky-Moriya coupling (allowed by
symmetry).  These materials exhibit band-like, large-polaron
transport with mobilities $\sim 1$~cm$^2$/V~s at $T\geq T_N$
\cite{CohnPolarons,ChiorescuMR}, which contrasts with the small
polaronic character of the paramagnetic phase of hole-doped
manganites \cite{HoleDoped}.

Figure~\ref{RofT} shows resistivity data (in zero magnetic field),
plotted versus inverse temperature, illustrating the sensitivity
of the charge transport in these materials to applied current ($I$)
at low $T$ where impurity conduction is predominant.  Several
curves for each specimen are labelled by values of $I$; successive
curves represent current increases in steps of 1, 3, 5 per decade.
This phenomenon is the central focus of this work.

The high-$T$ resistivity has an activated form, $\rho\propto\exp{(\Delta/k_BT)}$,
with $\Delta=86$~meV and 25~meV for CMO and SMO, respectively (dashed lines,
Fig.~\ref{RofT}).  The Hall coefficients ($R_H$) measured at the
lowest currents (squares and diamonds, Fig.~\ref{RofT}) follow
$\rho(T)$ in their temperature variation, consistent with
thermal activation of electrons from donor levels \cite{Pickett} to the
conduction band and a weakly $T$-dependent mobility.
Both $\rho$ and $R_H$ become weakly $T$ dependent in the impurity conduction regime at
the lowest $T$, but $R_H$ does not exhibit the maximum characteristic of a
transition to ``metallic'' impurity-band conduction.  This suggests that only a
small fraction of bound carriers at low-$T$ are sufficiently
mobile to contribute to $R_H$.  The conductivity ($\sigma$) extrapolates to zero as $T\to 0$ for the
lowest current (dashed line in inset, Fig.~\ref{RofT}), but $\sigma(T\to 0)$ for higher currents
is finite, indicating that the material is not strictly insulating.
\begin{figure}
\includegraphics[width =2.3in]{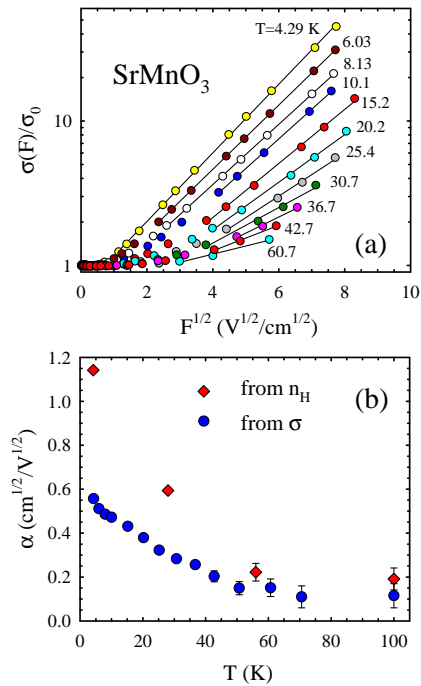}%
\vglue -.1in \caption{(Color online) (a) Semilog plot of
$\sigma(F)/\sigma_0$ {\it vs.} $F^{1/2}$ for SMO at various
temperatures. Solid lines are linear fits. (b) Slopes, $\alpha$,
determined from (a) (solid circles) and from a similar analysis of
the Hall density (Fig.~\ref{NsqrtF}).} \label{RofE} \vglue -.15in
\end{figure}

Isothermal measurements of $\sigma$ {\it
vs.}~applied current density ($J$) indicate that in the non-Ohmic
regime $\sigma$ scales with the transport electric field
($F\equiv\rho J$) as, $\sigma(F)/\sigma_0\propto
\exp(\alpha\sqrt{F})$, where $\sigma_0\equiv\sigma(I=5~nA)$. This
is shown for SMO in Fig.~\ref{RofE}~(a).  The temperature
dependence of the parameter $\alpha$, determined from
least-squares fits [solid lines, Fig.~\ref{RofE}~(a)], is plotted
in Fig.~\ref{RofE}~(b).  This dependence of $\sigma$ on $F$ is
that prescribed by Poole-Frenkel field-assisted ionization
\cite{Frenkel,PooleFrenkel} of carriers bound to donors, and is
discussed in more detail below.  Note that $\alpha$ was found to
be independent of applied magnetic field up to 9~T in spite of a
modest magneto-resistivity, e.g. $\Delta\rho/\rho|_{9T}\simeq
-0.13$ and $-0.04$ for $I=5$~nA and 0.5~mA, respectively, at
$T=28$~K for SMO.  Heating of the sample can be ruled out as the
cause of the $F$ dependence because specimen temperature was
monitored directly with a thermocouple, was limited to $\lesssim
2-3$~K at the highest current for each temperature, and was
corrected for by interpolation on the $\rho(T)$ curves at fixed
$I$.  Qualitatively similar results have been observed for these
same specimens after annealing to vary their oxygen content, and
for polycrystalline CMO and Ca$_{0.75}$Sr$_{0.25}$MnO$_3$
\cite{unpublished}.

\begin{figure}
\includegraphics[width =2.5in]{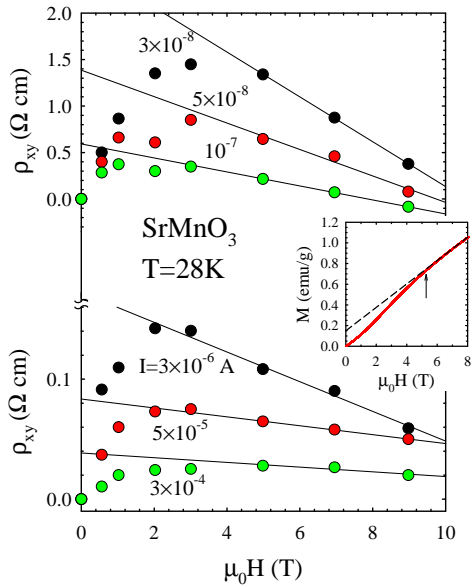}%
\caption{(Color online) Hall resistivity {\it vs.} magnetic field
for SMO at $T=28$~K.  Different curves are labeled by the applied
current.  The solid lines are linear fits at the highest fields
used to determine the normal Hall contribution. The inset shows
the magnetization at the same temperature (solid curve) and linear
fit to the high-field data (dashed line).}
\label{Hall} \vglue -.2in
\end{figure}

The current and magnetic field dependencies of the Hall resistivity, $\rho_{xy}$ (Fig.~\ref{Hall}),
provide further evidence that Poole-Frenkel ionization of trapped carriers
underlies the non-Ohmic behavior of $\sigma$.
This plot shows $\rho_{xy}$ {\it vs.} magnetic field for SMO at $T=28$~K
for several values of the applied current. With increasing field,
$\rho_{xy}$ for each current increases to a maximum near $\mu_0H\sim 3$~T, and becomes linear in field for
$\mu_0H\gtrsim 5$~T.
\noindent
This behavior indicates a sum of (positive) anomalous and (negative) normal Hall contributions,
consistent with the magnetization (inset, Fig.~\ref{Hall}) which exhibits a small FM contribution superposed with
the linear AF background.  The FM component saturates for $\mu_0H\gtrsim 5$~T, and the magnetization becomes linear
in field (dashed line) in the same field range as does $\rho_{xy}$.  The normal Hall coefficient was
determined from the high-field slopes, $R_H=d\rho_{xy}/d(\mu_0H)$ (solid lines, Fig.~\ref{Hall}).
Both $R_H$ and the anomalous contribution to $\rho_{xy}$ (intercept of solid lines) decrease systematically
with increasing current, signaling an increase in the mobile carrier density.
\begin{figure}
\includegraphics[width =4.2in]{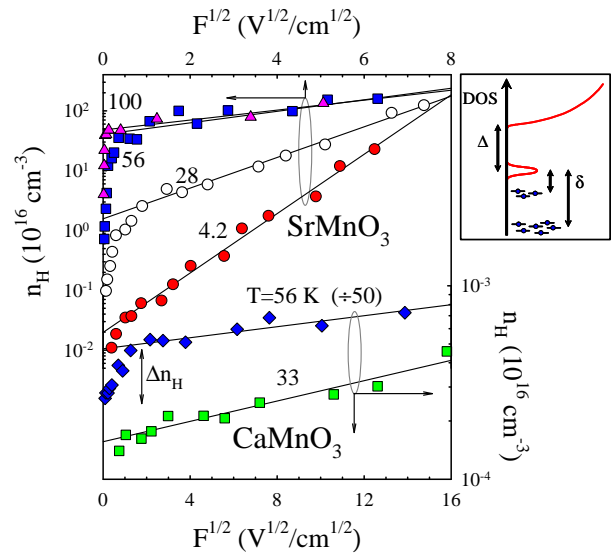}%
\vglue -.25in
\caption{(Color online) Semilog plots of Hall carrier density {\it
vs.} $F^{1/2}$ at different temperatures for CMO (lower abscissa,
right ordinate) and SMO (upper abscissa, left ordinate).  Solid
lines are linear fits. Inset: energy band scheme
implied by the temperature and electric-field dependence of $n_H$
(see text).} \label{NsqrtF}
\vglue -.2in
\end{figure}

Figure~\ref{NsqrtF} shows the Hall carrier density, $n_H\equiv
1/(R_He)$, plotted versus $F^{1/2}$ for CMO (lower abscissa, right ordinate)
and for SMO (upper abscissa, left ordinate) at several temperatures.
The ionization rate in the Poole-Frenkel model \cite{Frenkel,PooleFrenkel}
incorporates thermal activation as well as field-induced barrier lowering.  At low-$T$ where the majority of
electrons are trapped, the Hall density should be of the form,
$n_H(F)\approx N^0\exp[-(\delta/k_BT-\alpha\sqrt{F})]$, where $N^0$ is the density of neutral donors and
$\delta$ is the barrier height.  The scale for barrier lowering is set by $\alpha=\beta/k_BT$, where
$\beta=(Ze^3/\pi\epsilon_0\varepsilon_r)^{3/2}$ and $\varepsilon_r$
is an effective relative dielectric constant, often taken to be the high-frequency (optical) value.
The low-$T$ values of $\alpha$ determined from the slopes of linear
least-squares fits (solid lines, Fig.~\ref{NsqrtF}) imply $\beta\simeq 0.2$~meV cm$^{1/2}$/V$^{1/2}$ (CMO) and
0.4~meV cm$^{1/2}$/V$^{1/2}$ (SMO), in reasonable accord with the value 0.34~meV cm$^{1/2}$/V$^{1/2}$ (0.48~meV cm$^{1/2}$/V$^{1/2}$)
estimated for manganites using $Z=2$ (Z=1), for singly- (doubly-) occupied vacancies,
and $\varepsilon_r\simeq 5$ \cite{AlexandrovBratkovsky}.

A compelling
feature of the $n_H(F)$ data is that at each temperature, $n_H$ extrapolates toward
the value $n_H(300~{\rm K})$, the latter presumably reflecting
nearly full ionization of donors to the conduction band.  This implies that all carriers
are bound in Coulomb potentials in the ground state, and thus argues strongly against the existence of
self-trapped magnetic polarons in this material.  The $n_H(F)$ data for CMO show a much weaker
field effect, consistent with the smaller influence of $F$ on the resistivity (Fig.~\ref{RofT}).
The difference in the magnitude of the field effect for the two compounds implies a
difference in the potential barrier for donor ionization, $\delta$.
Applying the expression for $n_H$ above to the low-T $n_H$ data,
with $N^0=N_D-N_A\approx n_H(300~{\rm K})$, yields $\delta\simeq 41$~meV and 3.5~meV for CMO and SMO, respectively.
We conclude that electrons are not ionized to the conduction band (requiring energy $\sim\Delta$), but are
rather excited to a band of more mobile impurity states responsible for the low-$T$ conduction, as
depicted in the inset of Fig.~\ref{NsqrtF}.

Differing local environments, e.g. associated with vacancy
clusters or vacancy-acceptor pairs \cite{EfrosShklovskii}, are
expected to result in multiple bound-state energies. With
increasing $T$ electrons with larger binding energies are promoted
to the impurity band and rendered mobile in applied field $F$ so
that the $n_H(F,T)$ data provide for impurity-level spectroscopy.
The $n_H(F)$ curves for SMO at 28~K, 56~K, and 100~K imply
$\delta\approx 14\pm 3$~meV, indicating that there are principally
two such bound-state energies (inset, Fig.~\ref{NsqrtF}). The
larger values of $\Delta$ and $\delta$ for CMO suggest an intrinsic origin
(e.g. the buckled Mn-O-Mn bond), but compensation may also play a role.

Two aspects of the data indicate a strong electric-field dependence of the Hall mobility.
A low-field increase in the carrier density, designated $\Delta n_H$ in Fig.~\ref{NsqrtF} (vertical arrows),
comes to predominate in the total field effect as $T$ increases.  We attribute it to carriers
thermally excited to the impurity band where they are more loosely bound and mobile at low fields.
This field regime corresponds to {\it Ohmic\/} behavior in $\sigma$ (Fig.~\ref{RofE}), and thus implies
$\mu_H\propto n_H^{-1}$.  At higher fields within the Poole-Frenkel ionization regime, the values of $\alpha$
for SMO determined from $\sigma(F)$ are substantially lower than
those determined from $n_H(F)$ [Fig.~\ref{RofE}~(b)], by an amount that grows with decreasing
$T$ at $T\lesssim 60~{\rm K}\approx \delta$.  At 4.2~K, $\alpha_{n_H}\approx 2\alpha_{\sigma}$,
implying $\mu_H\propto n_H^{-1/2}$.  These observations suggest interesting correlation effects in the carrier dynamics.

It is likely that transport in the impurity band involves
next-nearest-neighbor Mn $e_g$ orbitals given that
nearest-neighbor hopping in the G-type AF lattice is strongly
inhibited by Hund's rule.  This band may involve excited impurity
states since little overlap is to be expected between bound states
with a mean spacing $d\approx 2[3/4\pi(N_D-N_A)]^{1/3}\sim
60-80~{\rm \AA}$, and radius $\lesssim 8-10~{\rm \AA}$ (estimated
from the donor polarizability \cite{CSMOeps}).

Though self-trapped MPs appear to be ruled out by our data, the
observation that a FM contribution to the magnetization in SMO
develops only below $\sim$~80~K \cite{Chmaissem}, is consistent
with bound MP formation for $k_BT\ll\delta$.  It is plausible that
MP de-trapping in electric field locally depolarizes a number of
Mn spins. The impurity-band carrier mobility may thus be
influenced by changes in both magnetic and Coulomb interactions
associated with de-trapping.

In summary, transport measurements in G-type antiferromagnetic
perovskite manganites indicate the presence of a mobile band of
impurity states to which electrons, bound at energy $\delta$
below, are excited through barrier-lowering in weak electric
fields. The data argue against the formation of self-trapped MPs
as predicted by theory, but are consistent with bound MPs that
become stable only for $k_BT\ll\delta$. Such an impurity band,
derived from states involving oxygen vacancies, is likely common
to related oxides where similar features in the transport might be
observed.  Such measurements can be employed as a tool for
impurity-level spectroscopy and for studying correlated electrons
with a tunable mobile carrier density.

The authors acknowledge helpful comments from S.~Satpathy. This material is based upon work supported by the National
Science Foundation under grants DMR-0072276 (Univ.~Miami) and DMR-0504769 (Mont. St. Univ.), and the Research Corporation (Univ.~Miami).


%
\vfill

\end{document}